\documentclass[prl,twocolumn,amsmath,amssymb,bibnotes,aps,longbibliography,superscriptaddress]{revtex4-1}
\usepackage{physics}
\usepackage{graphicx}
\usepackage{times}
\usepackage{xcolor}
\usepackage{ulem}
\usepackage{extarrows}
\usepackage[hmargin=1.7cm,vmargin=2.65cm]{geometry}
\usepackage{hyperref}

\definecolor{link}{RGB}{45,48,146}

\hypersetup{
pdftitle={A Quantum-Logic Gate between Distant Quantum-Network Modules},
pdfauthor={Severin Daiss, Stefan Langenfeld, Stephan Welte, Emanuele Distante, Philip Thomas, Lukas Hartung, Olivier Morin, Gerhard Rempe},
colorlinks=true, linkcolor=[RGB]{45,48,146}, citecolor=[RGB]{45,48,146}, filecolor=[RGB]{45,48,146}, urlcolor=[RGB]{45,48,146}}  
\renewcommand{\emph}{}

\newcommand{\be}{\begin{equation}}
\newcommand{\ee}{\end{equation}}

\begin{document} 
\title{A Quantum-Logic Gate between Distant Quantum-Network Modules}

\author{Severin Daiss}
\email{severin.daiss@mpq.mpg.de}
\affiliation{Max-Planck-Institut f\"ur Quantenoptik, Hans-Kopfermann-Strasse 1, 85748 Garching, Germany}
\author{\hspace{-.4em}\textcolor{link}{\normalfont\textsuperscript{, $\dagger$}}\hspace{.4em}Stefan Langenfeld}
\thanks{These authors contributed equally.}
\affiliation{Max-Planck-Institut f\"ur Quantenoptik, Hans-Kopfermann-Strasse 1, 85748 Garching, Germany}
\author{Stephan Welte}
\affiliation{Max-Planck-Institut f\"ur Quantenoptik, Hans-Kopfermann-Strasse 1, 85748 Garching, Germany}
\author{Emanuele Distante}
\affiliation{Max-Planck-Institut f\"ur Quantenoptik, Hans-Kopfermann-Strasse 1, 85748 Garching, Germany}
\affiliation{ICFO-Institut de Ciencies Fotoniques, The Barcelona Institute of Science and Technology, Mediterranean Technology Park, 08860 Castelldefels (Barcelona), Spain}
\author{Philip Thomas}
\affiliation{Max-Planck-Institut f\"ur Quantenoptik, Hans-Kopfermann-Strasse 1, 85748 Garching, Germany}
\author{Lukas Hartung}
\affiliation{Max-Planck-Institut f\"ur Quantenoptik, Hans-Kopfermann-Strasse 1, 85748 Garching, Germany}
\author{Olivier Morin}
\affiliation{Max-Planck-Institut f\"ur Quantenoptik, Hans-Kopfermann-Strasse 1, 85748 Garching, Germany}
\author{Gerhard~Rempe}
\affiliation{Max-Planck-Institut f\"ur Quantenoptik, Hans-Kopfermann-Strasse 1, 85748 Garching, Germany}

\begin{abstract}\noindent
The big challenge in quantum computing is to realize scalable multi-qubit systems with cross-talk free addressability and efficient coupling of arbitrarily selected qubits. Quantum networks promise a solution by integrating smaller qubit modules to a larger computing cluster. Such a distributed architecture, however, requires the capability to execute quantum-logic gates between distant qubits. Here we experimentally realize such a gate over a distance of 60m. We employ an ancillary photon that we successively reflect from two remote qubit modules, followed by a heralding photon detection which triggers a final qubit rotation. 
We use the gate for remote entanglement creation of all four Bell states. Our non-local quantum-logic gate could be extended both to multiple qubits and many modules for a tailor-made multi-qubit computing register.\\ 
\end{abstract}
\maketitle 

\noindent Quantum computers promise the capability to solve problems that cannot be dealt with by classical machines \cite{Feynman1982}. Despite recent progress with local qubit arrays \cite{Levine2019,Arute2019,Wright2019}, foreseeable implementations of quantum computers are limited in the number of individually controllable and mutually coupleable qubits held in one device due to residual cross-talk, coupling constraints, and restricted space \cite{Ladd2010}, to mention just three challenges. A solution to scale up the computational power is to embed the qubits into a non-local quantum network that, in essence, forms a large quantum machine \cite{Monroe2014}. This enables a modular approach to quantum computing, where small and spatially separated processing modules with accessible qubits can be arbitrarily connected with quantum links to carry out more complex joint calculations \cite{Kimble2008,Ladd2010}.

First elementary and fully coherent quantum networks with distant qubits have already demonstrated their ability to transfer quantum states or to create entanglement between distant quantum nodes \cite{Reiserer2015,Pirandola2015,Wehner2018}. This has been achieved with optical fibers that can carry photonic qubits, typically between different parts of a building or between nearby buildings. Combined with the small size of the modules containing the qubits, this is a key prerequisite for a distributed quantum computer with a potentially huge number of individually addressable qubits. For example, our modules are not much larger than a few $cm^3$ and can hold multiple independent atomic qubits \cite{Langenfeld2020}. Together with shared laser systems and single-atom reservoirs for uninterrupted operation \cite{Endres2016}, the number of connected modules and qubits can be straightforwardly scaled up. Replacing our free-space-optics modules with miniaturized cavity technology \cite{Thompson2013} could increase this number even further.

A mandatory step to extend state-of-the-art networks to a distributed quantum computing architecture is the ability to perform non-local two-qubit gates between arbitrary network modules. Towards this goal, gate teleportation has been proposed \cite{Gottesman1999}. This technique constitutes a generalization of teleportation and allows the construction of a quantum computer using previously shared entanglement, local operations and classical communication \cite{Eisert2000}. The general working principle has first been demonstrated with photonic systems based on linear optical quantum computing \cite{Huang2004,Gao2010} and, more recently, in isolated setups with material qubits like superconductors in a single cryostat \cite{Chou2018} and ions in a single Paul trap \cite{Wan2019}.

\begin{figure*}[t]
\begin{center}
\includegraphics[width=1.3\columnwidth]{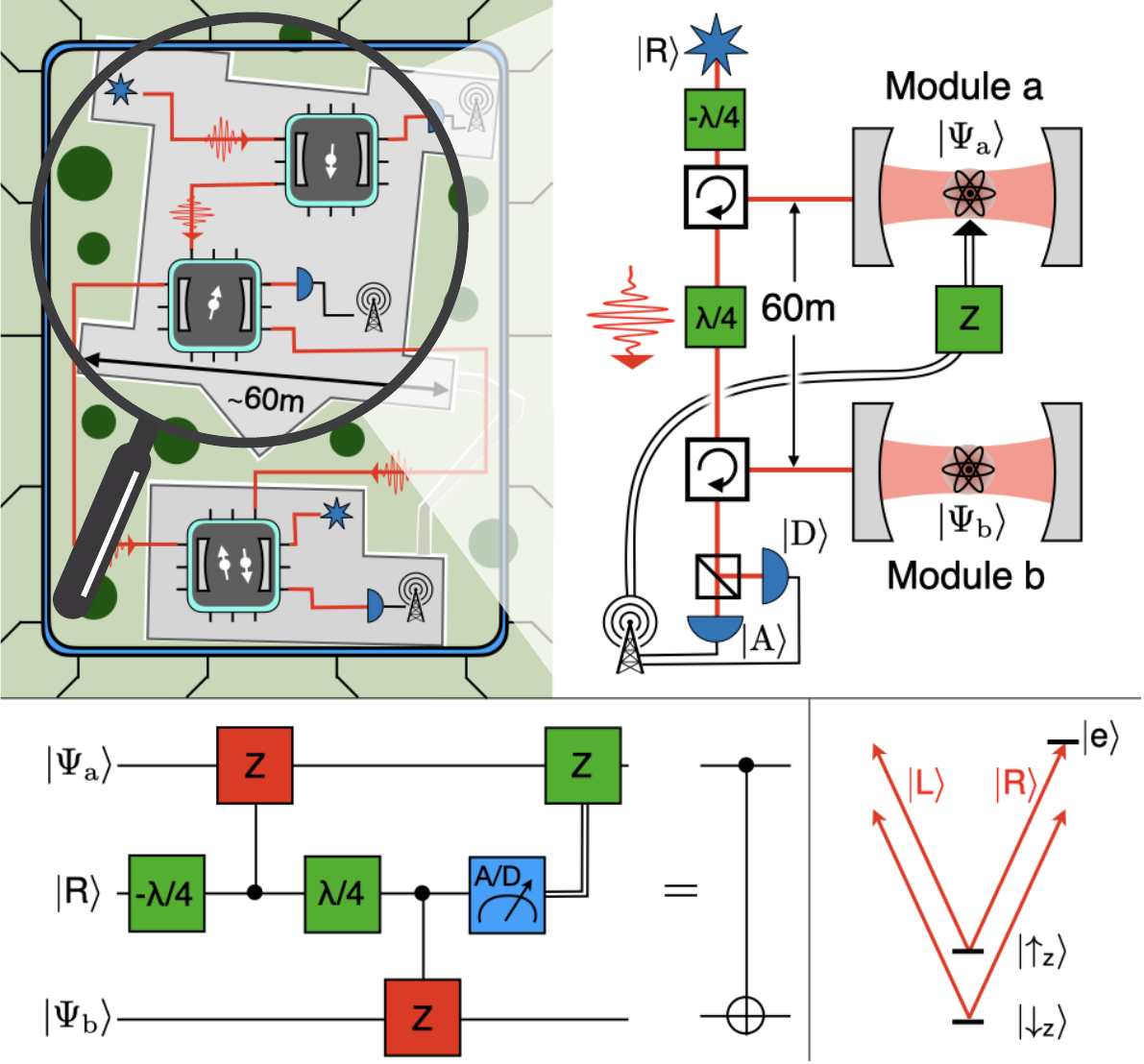}
\caption{\textbf{A quantum-logic gate between two distant qubit modules in a network.} (\textbf{A}) An elementary computing cluster formed by distributed qubit modules linked with optical fibers (red lines). The modules are connected to light sources (blue stars) and single-photon detectors (blue half discs) which feed classical communication channels (black antennas). (\textbf{B}) Enlarged view on our two qubit modules. They consist of an atom-cavity system each and are linked using an optical fiber and circulators. An ancilla photon is sent from a light source via the two modules to a polarization detector. The measurement result is communicated via a classical channel (double black line) to the first module for a conditional state rotation. (\textbf{C}) The quantum circuit diagram of our protocol which realizes the non-local quantum gate between the states $\ket{\Psi_\text{a}}$ and $\ket{\Psi_\text{b}}$ of the atomic qubits. (\textbf{D}) The relevant level scheme of the used atomic species $^{87}$Rb. }
\label{fig:principle}
\end{center}
\end{figure*}
Here we realize a non-local, universal quantum gate between two distant and completely independent quantum modules using a radically different and conceptually simple approach. Our scheme relies on a single photon as a flying ancilla qubit that we reflect successively from two distant network modules \cite{Duan2005,Lin2006}. They are connected by a $60\,\mathrm{m}$ fiber link and each contains a stationary qubit. The photon reflections establish two passive and in principle deterministic local gates between the photon and the qubits in the modules \cite{Reiserer2015}. Combined with a final measurement of the photon and feedback on the stationary qubit, our protocol implements a heralded, non-local quantum controlled-NOT (CNOT) gate. 
It has the advantage that the network modules are always ready for a gate operation and that a typically fragile entangled state does not need to be prepared and held available but is generated on the fly. The scheme is related to a local photon-photon gate protocol that was recently realized with the role of material and light qubits interchanged \cite{Hacker2016}.
The gate operation can be readily incorporated in a larger modularized quantum computer that is distributed over several buildings (Fig. \ref{fig:principle}A).

Our setup is schematically shown in Fig. \ref{fig:principle}B. It comprises two modules in separate laboratories each made up of a single $^{87}$Rb atom trapped at the center of an optical cavity with a finesse of 60\,000. We employ a light source to send photons to the modules and single-photon detectors register successful photon transmissions through the system. The resonators are single-sided and light interacting with them leaves predominantly through the incoming channel and is reflected. Two ground states of each atom $\ket{\uparrow_z}=\ket{5^2S_{1/2},F=2,m_F=2}$ and  $\ket{\downarrow_z}=\ket{5^2S_{1/2},F=1,m_F=1}$ serve as the qubit basis. We use a pair of Raman lasers to perform qubit rotations between them. The cavities are tuned to be resonant with the atomic transition $\ket{\uparrow_z}\leftrightarrow\ket{e}=\ket{5^2P_{3/2},F=3,m_F=3}$ (Fig. \ref{fig:principle}D). Both systems are in the strong coupling regime, thus the coherent atom-light interaction in the resonator modes have higher rates than the respective decays of the intra-cavity field and the atomic excitation \cite{Supplement}. Due to tailored light shifts on the energy levels of the atom, right-circular polarization $\ket{R}$ couples to an atomic transition, whereas left-circular polarization $\ket{L}$ does not. For an atom in $\ket{\uparrow_z}$ and coupling light in $\ket{R}$, this leads to a shift in the resonance frequency of the atom-cavity system \cite{Reiserer2015}. In this situation, an impinging photon resonant with the empty resonators cannot enter the cavities and is directly reflected. Only for an atom in the non-coupling state $\ket{\downarrow_z}$ or for left-circularly polarized light $\ket{L}$, the photon enters the resonator before leaving again through the incoupling mirror. In this process, it acquires a relative $\pi$ phase shift compared to directly reflected light, constituting a quantum Z gate between the photonic and the stationary qubit \cite{Reiserer2015}. For an incoming linear polarization $\ket{A}=1/\sqrt{2}\left(i\ket{R}+\ket{L}\right)$, this relative phase shift between $\ket{R}$ and $\ket{L}$ flips the polarization to an orthogonal state $\ket{D}=1/(\sqrt{2}i)\left(i\ket{R}-\ket{L}\right)$ if the atom is in a coupling state $\ket{\uparrow_z}$. As the polarization remains unchanged for an atom in $\ket{\downarrow_z}$, the reflection can entangle the photonic qubit to the atomic state and it thus forms the basic building block of our non-local quantum gate. 

The circuit diagram for our protocol is shown in Fig. \ref{fig:principle}C. We start with a right circular polarized photon $\ket{R}$ with a duration of $1\,\mathrm{\mu s}$ (full width at half maximum) and use a wave plate to convert it to the linear polarization $\ket{A}$. In practice, we approximate the single photon by a weak coherent laser pulse with an average photon number of $\bar{n}=0.07\pm0.01$. We tune it to be resonant with the empty resonators and reflect it from the first cavity, thereby entangling the photon polarization and the atomic spin.
The light is subsequently sent through a quarter wave plate and a $60\,\mathrm{m}$ long optical single-mode fiber to the other module located $21\,\mathrm{m}$ away. Passing a circulator, the photon is reflected from the second system resulting in an atom-atom-photon entangled state. Finally, the light is sent to a polarization-detection setup measuring in the linear basis of $\ket{A}/\ket{D}$. The detection of the photon acts as a herald of a successful gate operation and triggers a rotation of the state of the first qubit conditioned on the measurement outcome. The resulting overall protocol realizes a non-local quantum gate. In the basis of $\ket{\uparrow_z}/\ket{\downarrow_z}$ for the first atom and in the superposition basis of $\ket{\uparrow_x}=\frac{1}{\sqrt{2}}\left(\ket{\uparrow_z}+\ket{\downarrow_z}\right)$ and $\ket{\downarrow_x}=\frac{1}{\sqrt{2}}\left(\ket{\uparrow_z}-\ket{\downarrow_z}\right)$ for the second atom this gate acts as a CNOT 
\be \begin{split}\ket{\uparrow_z\uparrow_x}& \longrightarrow \ket{\uparrow_z\uparrow_x}\\
\ket{\uparrow_z\downarrow_x}& \longrightarrow \ket{\uparrow_z\downarrow_x}\\
\ket{\downarrow_z\uparrow_x}& \longrightarrow \ket{\downarrow_z\downarrow_x}\\
\ket{\downarrow_z\downarrow_x}& \longrightarrow \ket{\downarrow_z\uparrow_x}.
\end{split}\ee
Here we adopt a common notation for the two distant qubits with the first (second) state belonging to the qubit in the first (second) module. A detailed derivation of our gate protocol and the corresponding truth table is given in the supplement \cite{Supplement}.

\begin{center}
\begin{figure}
\includegraphics[width=\columnwidth]{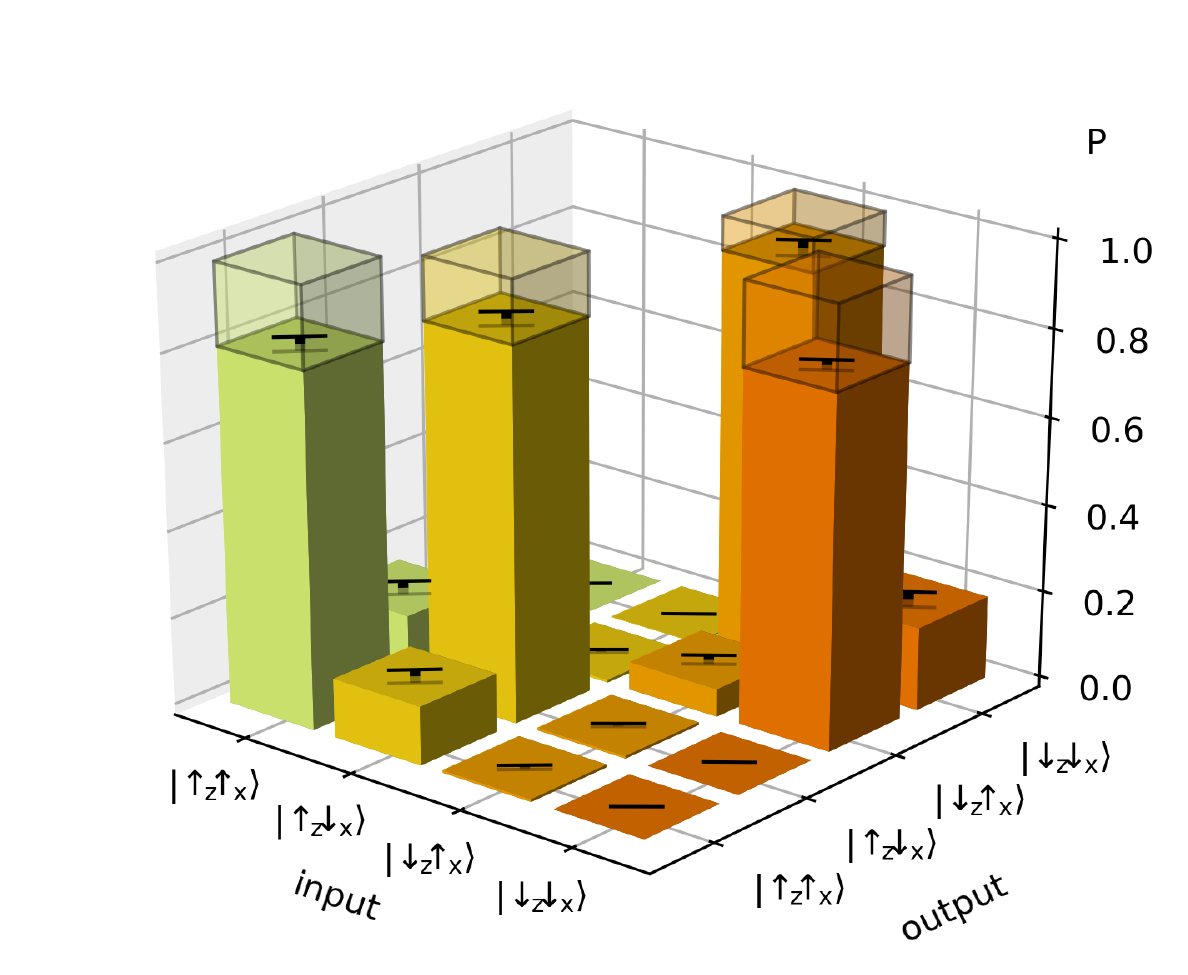}
\caption{\textbf{Truth table of the non-local protocol.} The diagram shows the probability $P$ to measure a certain gate output state for the different input states $\ket{\uparrow_z\uparrow_x}$, $\ket{\uparrow_z\downarrow_x}$, $\ket{\downarrow_z\uparrow_x}$ and $\ket{\downarrow_z\downarrow_x}$. In this basis, our protocol acts as a CNOT gate. Errors are given as black ranges and indicate the standard deviations in our measurement. The expected output states for an ideal CNOT gate are shown as light shaded bars.}
\label{fig:truth table}
\end{figure}
\end{center}

We probe our gate with all combinations of $\ket{\uparrow_z}/\ket{\downarrow_z}$ for the first and $\ket{\uparrow_x}/\ket{\downarrow_x}$ for the second qubit as input states and measure the resulting atomic states in the $z$ and $x$ bases, respectively. For each input combination, we register about $500$ heralding clicks. This truth table of our operation is given in Fig. \ref{fig:truth table}. We achieve a fidelity of $(85.1\pm0.8)\%$ with the populations expected for an ideal quantum CNOT gate.

\begin{center}
\begin{figure}
\includegraphics[width=\columnwidth]{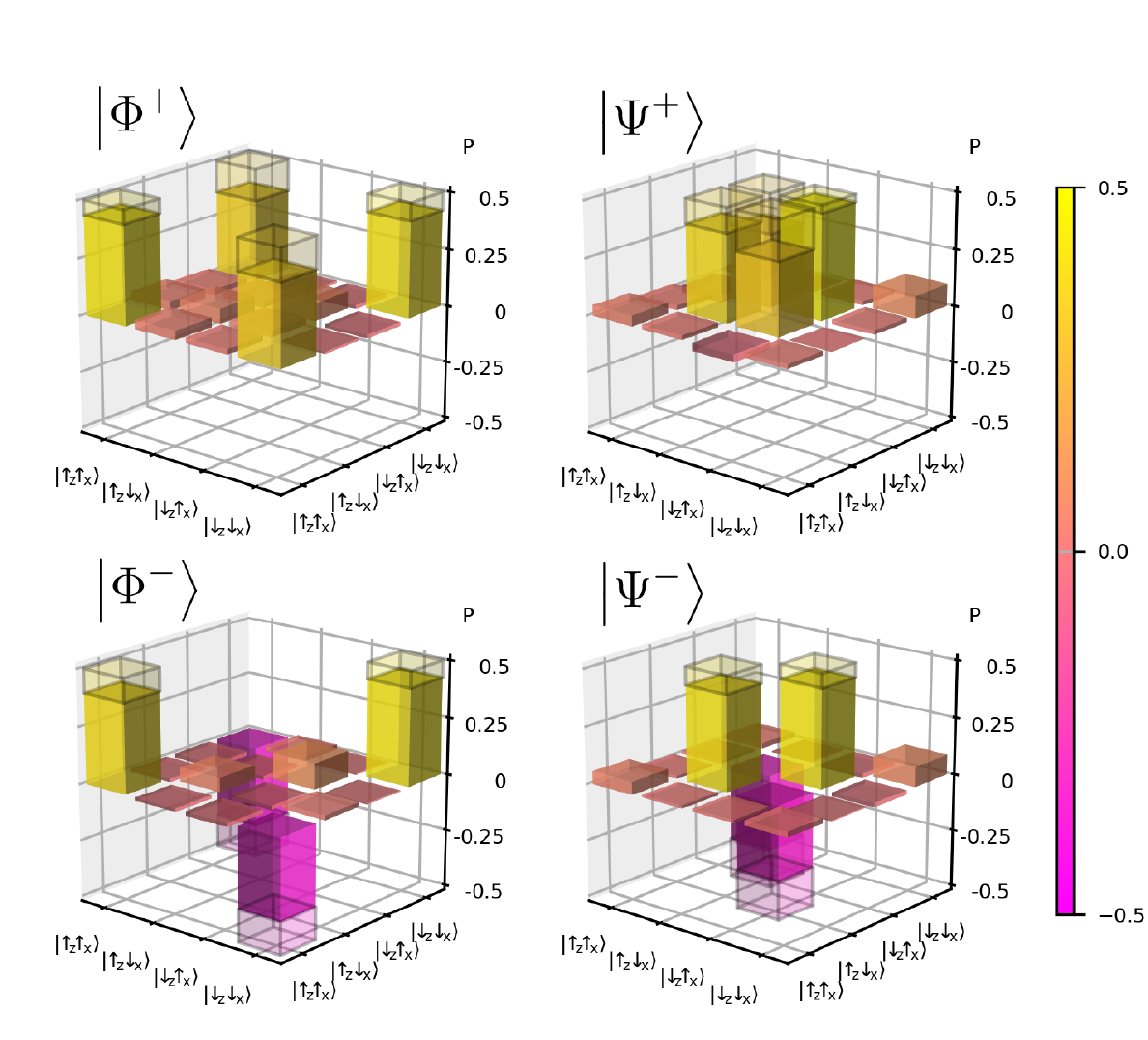}
\caption{\textbf{Density matrices for the produced Bell states.} The real part of the reconstructed density matrices are shown. The ideal values are indicated as shaded bars in the plots.}
\label{fig:bell}
\end{figure}
\end{center}

To rule out possible non-zero phases in the unitary underlying the truth table (for a more detailed discussion see supplement \cite{Supplement}) and to confirm the quantum nature of our non-local gate, we furthermore use it to entangle the two qubits. For this purpose, we initialize each atom either in $\ket{\uparrow_x}$ or $\ket{\downarrow_x}$, execute our protocol and perform a full state tomography on the resulting two-qubit state by measuring both atoms in random combinations of detection bases \cite{Altepeter2004}. Depending on the initial state, we generate all four maximally entangled Bell states. For each of them, we collect about $3000$ heralding photon detections. The results of this measurement are displayed in Table \ref{tab:bell_states} and Fig. \ref{fig:bell}. The table gives the final fidelities with the ideal Bell states and the figure displays the real parts of the reconstructed density matrices for each input state. The imaginary residues are given in the supplement \cite{Supplement}. We achieve an average overlap fidelity with ideal Bell states of $(76.6\pm1.0)\%$.
\begin{table}
\begin{center}
\begin{tabular}{|c|c|c|}%
\hline
Input state & Output state & Fidelity\\
\hline
$\ket{\uparrow_{x}\uparrow_{x}}$ & $\frac{1}{\sqrt{2}}\left(\ket{\uparrow_z\uparrow_x}+{\downarrow_z\downarrow_x}\right)=\ket{\Phi^+}$ & $(78.8\pm 2.0)\%$ \\
$\ket{\uparrow_{x}\downarrow_{x}}$ & $\frac{1}{\sqrt{2}}\left(\ket{\uparrow_z\downarrow_x}+{\downarrow_z\uparrow_x}\right)=\ket{\Psi^+}$ & $(75.1\pm 2.0)\%$ \\
$\ket{\downarrow_{x}\uparrow_{x}}$ & $\frac{1}{\sqrt{2}}\left(\ket{\uparrow_z\uparrow_x}-{\downarrow_z\downarrow_x}\right)=\ket{\Phi^-}$ & $(76.8\pm 2.0)\%$ \\
$\ket{\downarrow_{x}\downarrow_{x}}$ & $\frac{1}{\sqrt{2}}\left(\ket{\uparrow_z\downarrow_x}-{\downarrow_z\uparrow_x}\right)=\ket{\Psi^-}$ & $(75.8\pm 2.0)\%$ \\
\hline
\end{tabular}%
\caption{\label{tab:bell_states}
\textbf{Bell-state generation results.} List of input and output states of the generation of the Bell states with our non-local CNOT gate protocol up to a global phase. The last column gives the measured fidelities with the expected final state.
}
\end{center}
\end{table}

The whole protocol is performed at a rate of $1\,\mathrm{kHz}$. The gate itself takes $22\,\mu\mathrm{s}$, limited by the time needed for single-qubit rotations. While in our implementation this is still slow compared to state-of-the-art quantum computers, the gate serves as a special connecting gate and might not need to run with the same high repetition rate and speed.

The success probability to detect the heralding photon is relatively small, $0.006$, with a major limitation coming from our use of a weak coherent laser pulse with a vacuum contribution of $93\%$.
Furthermore, the efficiency is lowered by the cavity reflectivities (60\% and 55\%), the link between the two systems including optical elements, a fiber coupling, photon loss in the fiber, and a circulator transmission (52\% in total) and the detection efficiency after the second module including again a circulator transmission (50\% in total). The only intrinsic limitation to the success probability is the photon transmission in the $60\,\mathrm{m}$ fiber link (95\% for a wavelength of $780\,\mathrm{nm}$).
\begin{table*}
\begin{center}
\bgroup
\def\arraystretch{1.3}
\begin{tabular}{|c|c|c|}%
\hline
Imperfection & Effect on truth table & Effect on Bell states \\
\hline
Weak coherent states and losses & $(2.3^{+0.3}_{-0.3})\%$ & $(3.8^{+0.5}_{-0.4})\%$\\
State-preparation and measurement & $(2.9^{+1.0}_{-1.0})\%$ & $(2.7^{+1.0}_{-1.0})\%$\\
Polarization effects & $(2.2^{+0.3}_{-0.5})\%$ & $(2.7^{+0.4}_{-0.6})\%$\\
Mode matching & $(2.0^{+1.6}_{-1.2})\%$ & $(2.4^{+2.5}_{-1.5})\%$\\
Atom-cavity detunings and lock widths & $(1.4^{+1.3}_{-0.7})\%$ & $(1.9^{+2.8}_{-1.0})\%$\\
Atomic decoherence & $(0.11^{+0.09}_{-0.03})\%$ & $(3.2^{+0.3}_{-0.2})\%$\\
\hline
\end{tabular}%
\egroup
\caption{\label{tab:imperfection}
\textbf{Effects of experimental imperfections.} List of major experimental imperfections in our implementation and their effective fidelity reduction on Bell-state production and truth table measurements derived from a separate simulation. Errors are estimated from the
fluctuations of the experimental parameters.
}
\end{center}
\end{table*}
The fidelity of our non-local quantum gate is currently limited by various experimental imperfections. We have identified the major effects and performed simulations to estimate their reduction of fidelity, both for the truth table and for the generation of entanglement. The results are listed in Table \ref{tab:imperfection}. These imperfections are not intrinsic to the protocol and are expected to be improved in future implementations.

One of the main contributions stems from our use of a weak coherent laser pulse that has a residual probability of a two-photon state impinging on either cavity. Such a contribution changes the relative $\pi$ phase in the atom-light state between a coupling and a non-coupling atom to a full $2 \pi$ phase, thereby nullifying the effect caused by the atom. The resulting infidelity can be readily vanquished by using a single photon, for example derived from a well-controlled atom-cavity system \cite{Reiserer2015}, as an ancilla between the modules.

Additional limitations come from our state preparation and measurement, the finite coherence time of the atoms, the transversal mode matching between fibers and cavities, the frequency stability of both resonators and the matching of polarizations between the different modules and the detection setup. Some of these effects have a smaller impact on the fidelity of the truth table for two reasons. First, in this measurement the qubit in the first module is prepared in an energy eigenstate and second, additional polarization effects and higher photon numbers do not affect the fidelity with a non-coupling atom in $\ket{\downarrow_z}$. Smaller errors in both measurements in the sub-percent range arise from dark counts of the detectors and the parameters of our cavities. A more detailed discussion of the imperfections is given in the supplement \cite{Supplement}. 
Overall, we expect to reduce theses sources of error with measures as using single photons, implementing faster single qubit rotations and better long-term stability of the relevant frequencies and polarizations.

In summary, we have demonstrated a non-local quantum gate between two separated qubit modules connected with a $60\,\mathrm{m}$ optical fiber.
While we implement this gate using single atoms coupled to an optical cavity, the general principle can be transferred to different platforms like quantum dots \cite{Lodahl2015} or impurity-vacancies in diamond \cite{Bhaskar2020}.
We are using weak coherent states to approximate a single photon, but better sources like atom-cavity systems could greatly improve both the efficiency and the fidelity of our implementation. Arguably most fascinating, however, is to extend our experiment to more than two modules and more than one qubit per module, including local two-qubit gates \cite{Welte2018}. This could facilitate in a controlled way a single-photon entangled multi-qubit array \cite{Cohen2018}, or a quantum repeater chain \cite{Uphoff2016}.

\bibliographystyle{bibstyle_dist}

\section*{Acknowledgments}
\noindent We thank B. Hacker, M. K\"{o}rber, and T. Nadolny for contributions during the early stage of the experiments. \textbf{Funding:} This work was supported by the Bundesministerium f\"{u}r Bildung und Forschung via the Verbund Q.Link.X (16KIS0870), by the Deutsche Forschungsgemeinschaft under Germany’s Excellence Strategy – EXC-2111 – 390814868, and by the European Union’s Horizon 2020 research and innovation programme via the project Quantum Internet Alliance (QIA, GA No. 820445). E.D. acknowledges support by the Cellex-ICFO-MPQ postdoctoral fellowship program. \textbf{Authors contributions:} All authors contributed to the experiment, the analysis of the results and the writing of the manuscript.
\textbf{Competing interests:} The authors declare that there are no competing financial interests. 
\textbf{Data and materials availability:} The data underlying the figures can be found in the zenodo repository at doi: 10.5281/zenodo.4284323. All other data needed to evaluate the conclusions in the paper are present in the paper or the Supplementary Materials.

\clearpage
\newpage
\section*{Supplementary Text}
\renewcommand{\thefigure}{S\arabic{figure}}
\setcounter{figure}{0}
\renewcommand{\thetable}{S\arabic{table}}
\setcounter{table}{0}

\subsection*{Experimental Setup}
\noindent In this work, we use two independent qubit modules that are physically separated by $21\text{m}$ and connect the respective laboratories using a $60\text{m}$ optical single-mode fiber as a network link. We employ piezoelectric fiber squeezers and reference beams with a fixed polarization to control the birefringence of the fiber link to act as a quarter wave plate and to compensate slow thermal drifts in regular time intervals \cite{Rosenfeld2008}. In front of the first module, we employ a $1.5\%$ reflecting, non-polarizing beam splitter as a circulator, in front of the second module we use a bidirectional option with an $80\%$ transmission. Each module contains a single-sided cavity with a $^{87}$Rb atom that strongly couples to the resonator and is trapped in an optical lattice. 
\begin{table*}
\begin{center}
\begin{tabular}{|c|c|c|}%
\hline
 & Module a & Module b \\
\hline
qubit-cavity coupling g & $2\pi\times7.6\text{MHz}$ & $2\pi\times7.6\text{MHz}$\\
cavity decay rate $\kappa$ & $2\pi\times2.5\text{MHz}$ & $2\pi\times2.8\text{MHz}$\\
cavity decay into outcoupling mode $\kappa_r$ & $2\pi\times2.3\text{MHz}$ & $2\pi\times2.4\text{MHz}$\\
atomic decay rate $\gamma$ & $2\pi\times3\text{MHz}$ & $2\pi\times3\text{MHz}$\\
\hline
\end{tabular}%
\caption{\label{tab:cavQEDparams}
\textbf{cQED parameters.} List of the relevant cQED parameters of the two modules used.
}
\end{center}
\end{table*}
Table \ref{tab:cavQEDparams} lists the relevant cavity quantum electrodynamics (cQED) parameters of the systems, where the qubit space is given by the states $\ket{\uparrow_z}=\ket{5^2S_{1/2},F=2,m_F=2}$ and  $\ket{\downarrow_z}=\ket{5^2S_{1/2},F=1,m_F=1}$ and the cavities are tuned to the atomic transition $\ket{\uparrow_z}\leftrightarrow\ket{e}=\ket{5^2P_{3/2},F=3,m_F=3}$. Here Module a (b) indicates the resonator having the first (second) interaction with the mediating photon.
Initially, the atoms are pumped to the state $\ket{\uparrow_z}$ during $200\mu\text{s}$ using right-circularly polarized light along the cavity axis. We can manipulate the atomic state at each module with a pair of Raman beams to perform rotations of the qubits. With an adjustment of the phase between the two beams, the rotations can be done around both the $x$-axis ($ \frac{1}{\sqrt{2} } \left( \ket{\uparrow_z}+\ket{\downarrow_z} \right)$) and the $y$-axis ($\frac{1}{\sqrt{2}}\left(\ket{\uparrow_z}+i\ket{\downarrow_z} \right)$). In the present experiment, a $\pi$ pulse is peformed within $8\mu s$ at both setups.
To detect the state of the atom at the end of an experiment, we impinge laser light resonant with the transition $\ket{\uparrow_z}\leftrightarrow\ket{e}$ from the side on the atom. For an atom in $\ket{\uparrow_z}$, photons are scattered into the cavity mode and leave through the in- and outcoupling mirror. Using an acousto-optical modulator as a path switch, the state detection light from Module a can be sent directly to a detection setup without using the quantum network link. For Module b, the light is sent to superconducting nanowire single-photon detectors using the circulator.

\subsection*{Protocol}
\subsubsection*{Definitions}
\noindent For the different qubits used in the protocol, we use the basis \be\ket{\uparrow_z}=\begin{pmatrix}1\\0\end{pmatrix},\hspace{1em}\ket{\downarrow_z}=\begin{pmatrix}0\\1\end{pmatrix}\ee for each atom and the basis \be\ket{R}=\begin{pmatrix}1\\0\end{pmatrix},\hspace{1em}\ket{L}=\begin{pmatrix}0\\1\end{pmatrix}\ee for the polarization. The other relevant polarization states are defined as
\be\begin{split}
&\ket{D}=\frac{1}{\sqrt{2}i}\left(i\ket{R}-\ket{L}\right),\\
&\ket{A}=\frac{1}{\sqrt{2}}\left(i\ket{R}+\ket{L}\right).\\
\end{split}\ee
The effect of a quarter wave plate in this basis is given as a transformation
\be T_{\lambda/4} = \frac{1}{\sqrt{2}}\begin{pmatrix}-i & 1 \\ 1 & -i\end{pmatrix},\hspace{1em}
T_{-\lambda/4} = \frac{1}{\sqrt{2}}\begin{pmatrix}i & 1 \\ 1 & i\label{quarterwave} \end{pmatrix}.\ee 
For the atomic state, the effect of $\pi$ and $\pi/2$ rotations around the x- and y-axes on our basis states are given by the transformation matrices
\be T_{\pi/2}^y = \frac{1}{\sqrt{2}}\begin{pmatrix}1 & -1 \\ 1 & 1 \end{pmatrix},\hspace{1em}
T_{\pi/2}^x = \frac{1}{\sqrt{2}}\begin{pmatrix}1 & -i \\ -i & 1 \end{pmatrix}\ee and
\be T_{\pi}^y = \begin{pmatrix}0 & -1 \\ 1 & 0 \end{pmatrix},\hspace{1em}
T_{\pi}^x = \begin{pmatrix}0 & -i \\ -i & 0 \end{pmatrix},\ee
where the angle is given by the subscript and the axis is indicated by the superscript.
\subsubsection*{Reflection from the Cavity}
\noindent In the following, the cavity is tuned to be on resonance with the transition $\ket{\uparrow_z}\longrightarrow\ket{e}$. Due to a normal mode splitting for the atom in the coupling state $\ket{\uparrow_z}$, light in the $\ket{R}$ polarization is blocked from entering the cavity and is directly reflected without any further phase shift. For $\ket{L}$ polarized light or the non-coupling state $\ket{\downarrow_z}$, the light enters the cavity and acquires a $\pi$ phase shift in the reflection process of leaving again through the incoupling mirror. The effect of a reflection from the cavity on the other relevant polarizations is given as \be\setlength{\jot}{9pt}\begin{split}\ket{\uparrow_z}\ket{A}\rightarrow & \frac{1}{\sqrt{2}}\ket{\uparrow_z}\left(i\ket{R}-\ket{L}\right)=i\ket{\uparrow_z}\ket{D},\\ \ket{\downarrow_z}\ket{A}\rightarrow & -\ket{\downarrow_z}\ket{A},\\
\ket{\uparrow_z}\ket{D}\rightarrow & \frac{1}{\sqrt{2}}\ket{\uparrow_z}\left(\ket{R}-i\ket{L}\right)=-i\ket{\uparrow_z}\ket{A} \\
\ket{\downarrow_z}\ket{D}\rightarrow & -\ket{\downarrow_z}\ket{D}.\label{eq:cav2}\end{split}\ee 
\subsubsection*{Full Protocol of the Non-Local Quantum Gate}
\noindent To detail our non-local gate protocol, we consider its effect on a combined atomic state
\be \ket\Psi = \alpha\ket{\uparrow_z\uparrow_z}+\beta\ket{\uparrow_z\downarrow_z}+\delta\ket{\downarrow_z\uparrow_z}+\xi\ket{\downarrow_z\downarrow_z}.\ee 
Here the first spin in each ket indicates the qubit state in the module from that the photon is first reflected, whereas the second spin corresponds to the qubit state in the second module. For brevity, we omit additional normalization factors in the following discussion. 

We start with a photon polarized in $\ket{R}$ and use a $-\lambda/4$ wave plate to convert it to $\ket{A}$. Using Eq. \ref{eq:cav2}, the first reflection yields the combined state
\be\begin{split} \ket\Psi =& i\alpha\ket{\uparrow_z\uparrow_z}\ket{D}+i\beta\ket{\uparrow_z\downarrow_z}\ket{D}\\&-\delta\ket{\downarrow_z\uparrow_z}\ket{A}-\xi\ket{\downarrow_z\downarrow_z}\ket{A}.\end{split}\ee
Being transmitted to the second setup, the light passes through a $\lambda/4$ wave plate, converting $\ket{A}\rightarrow\ket{R}$ and $\ket{D}\rightarrow\ket{L}$ (see Eq. \ref{quarterwave}). After the reflection from the second cavity, we have then
\be \begin{split}\ket\Psi =& -i\alpha\ket{\uparrow_z\uparrow_z}\ket{L}-i\beta\ket{\uparrow_z\downarrow_z}\ket{L}\\&-\delta\ket{\downarrow_z\uparrow_z}\ket{R}+\xi\ket{\downarrow_z\downarrow_z}\ket{R}.\label{tworefl}\end{split}\ee
Lastly, we detect the photon in the superposition basis of $\ket{A}/\ket{D}$. Rewriting the circular polarizations as $\ket{R}=\ket{D}-i\ket{A},\ket{L}=\ket{A}-i\ket{D}$, we get
\be\begin{split}\ket\Psi =&-(\alpha\ket{\uparrow_z\uparrow_z}+\beta\ket{\uparrow_z\downarrow_z}+\delta\ket{\downarrow_z\uparrow_z}-\xi\ket{\downarrow_z\downarrow_z})\ket{D}\\&-i(\alpha\ket{\uparrow_z\uparrow_z}+\beta\ket{\uparrow_z\downarrow_z}-\delta\ket{\downarrow_z\uparrow_z}+\xi\ket{\downarrow_z\downarrow_z})\ket{A}.\end{split}\ee
\noindent The photon-detection result is communicated to the first network module. For a detection in $\ket{A}$, there is an effective Z-gate feedback applied to the first atom by a combination of a $\pi$ pulse around the $y$-axis followed by a $\pi$ pulse around the $x$-axis:
\begin{widetext}
\be iT_{\pi}^{x,1}T_{\pi}^{y,1}\left(\alpha\ket{\uparrow_z\uparrow_z}+\beta\ket{\uparrow_z\downarrow_z}-\delta\ket{\downarrow_z\uparrow_z}+\xi\ket{\downarrow_z\downarrow_z}\right)=\alpha\ket{\uparrow_z\uparrow_z}+\beta\ket{\uparrow_z\downarrow_z}+\delta\ket{\downarrow_z\uparrow_z}-\xi\ket{\downarrow_z\downarrow_z}.\ee\end{widetext}
The superscript $1$ indicates that the rotation is only applied to the first atom. For a detection in $\ket{D}$, no feedback is applied.
In total, this realizes the final state
\be \begin{split}\ket\Psi = -(\alpha\ket{\uparrow_z\uparrow_z}+\beta\ket{\uparrow_z\downarrow_z}+\delta\ket{\downarrow_z\uparrow_z}-\xi\ket{\downarrow_z\downarrow_z}).\end{split}\ee
and thus a fully controlled quantum-PHASE gate. Combining this with two Hadamard gates or considering the effect of the gate on a superposition basis as $\ket{\uparrow_x}=\frac{1}{\sqrt{2}}\left(\ket{\uparrow_z}+\ket{\downarrow_z}\right)$ and $\ket{\downarrow_x}=\frac{1}{\sqrt{2}}\left(\ket{\uparrow_z}-\ket{\downarrow_z}\right)$, we get a quantum CNOT gate described by
\be \label{eq:truthtable}\begin{split}\ket{\uparrow_z\uparrow_x}& \longrightarrow \ket{\uparrow_z\uparrow_x}\\
\ket{\uparrow_z\downarrow_x}& \longrightarrow \ket{\uparrow_z\downarrow_x}\\
\ket{\downarrow_z\uparrow_x}& \longrightarrow \ket{\downarrow_z\downarrow_x}\\
\ket{\downarrow_z\downarrow_x}& \longrightarrow \ket{\downarrow_z\uparrow_x}
\end{split}\ee
and the unitary \be
\hat{U}_\mathrm{CNOT}=\begin{pmatrix}
1 & 0 & 0 & 0 \\
0 & 1 & 0 & 0 \\
0 & 0 & 0 & 1 \\
0 & 0 & 1 & 0
\end{pmatrix}
\ee
acting on states in the basis $\ket{\uparrow_z\uparrow_x}$, $\ket{\uparrow_z\downarrow_x}$, $  \ket{\downarrow_z\uparrow_x}$, $\ket{\downarrow_z\downarrow_x}$.
\subsubsection*{Ruling Out Possible Non-Zero Phases in the Implemented Transformation}
\noindent The truth table for the CNOT gate is determined by measuring the output populations for all four input basis states. However, this does not yield information about possible non-zero phases in the experimental implementation. Using the basis states from above, the information obtained from the truth table of an ideal CNOT gate can in general be described by a matrix of the form
\be\label{eq:gate-phases}
\hat{U}_{\alpha,\beta,\gamma}=\begin{pmatrix}
1 & 0 & 0 & 0 \\
0 & \exp{i\alpha} & 0 & 0 \\
0 & 0 & 0 & \exp{i\beta} \\
0 & 0 & \exp{i\gamma} & 0
\end{pmatrix}\vspace{0.1cm}
\ee
with some phases $\alpha,\beta,\gamma$. The complex phase of the first column/row can be absorbed in a global phase and can be set to zero without loss of generality. To determine the phases in this unitary, additional measurements besides the truth table are needed.
Information about the realized phases can be derived from Bell-state production experiments. For example, acting with $\hat{U}_{\alpha,\beta,\gamma}$ on $\ket{\uparrow_x\uparrow_x}$ results in Eq. \ref{eq:phase1}.
\begin{widetext}\be \label{eq:phase1}\ket{\uparrow_x\uparrow_x}=\frac{1}{\sqrt{2}}\left(\ket{\uparrow_z\uparrow_x} + \ket{\downarrow_z\uparrow_x}\right)\xrightarrow{\makebox[2cm]{ $\hat{U}_{\alpha,\beta,\gamma}$ }}\frac{1}{\sqrt{2}}\left(\ket{\uparrow_z\uparrow_x} + \exp{i\gamma} \ket{\downarrow_z\downarrow_x}\right).\ee
\be \label{eq:phase2}\ket{\uparrow_x\downarrow_x}=\frac{1}{\sqrt{2}}\left(\ket{\uparrow_z\downarrow_x} + \ket{\downarrow_z\downarrow_x}\right)\\\xrightarrow{\makebox[2cm]{ $\hat{U}_{\alpha,\beta,\gamma}$ }}\frac{1}{\sqrt{2}}\left(\exp{i\alpha}\ket{\uparrow_z\downarrow_x} + \exp{i\beta} \ket{\downarrow_z\uparrow_x}\right).\ee
\be \label{eq:phase3}\ket{\uparrow_z\uparrow_z}=\frac{1}{\sqrt{2}}\left(\ket{\uparrow_z\uparrow_x} + \ket{\uparrow_z\downarrow_x}\right)\xrightarrow{\makebox[2cm]{ $\hat{U}_{\alpha,\beta,\gamma}$ }}\frac{1}{\sqrt{2}}\left(\ket{\uparrow_z\uparrow_x} + \exp{i\alpha} \ket{\uparrow_z\downarrow_x}\right).\ee\end{widetext}
If the experiment produces the Bell state $\ket{\Phi^+}=\left(\ket{\uparrow_z\uparrow_x} + \ket{\downarrow_z\downarrow_x}\right)/\sqrt{2}$, we conclude $\gamma=0$.
Similarly, $\hat{U}_{\alpha,\beta,\gamma}$ applied to $\ket{\uparrow_x\downarrow_x}$ yields Eq. \ref{eq:phase2}.
Here, if the experiment produces the Bell state $\ket{\Psi^+}=\left(\ket{\uparrow_z\downarrow_x} + \ket{\downarrow_z\uparrow_x}\right)/\sqrt{2}$, we conclude $\alpha=\beta$. These two conclusions leave only one relative phase, namely $\alpha$, unknown. 
To see the effect of a possibly non-zero phase $\alpha$, $\hat{U}_{\alpha,\beta,\gamma}$ is applied to the state $\ket{\uparrow_z\uparrow_z}$ to get the result of Eq. \ref{eq:phase3}.
If the experiment produces the state $\left(\ket{\uparrow_z\uparrow_x} + \ket{\uparrow_z\downarrow_x}\right)/\sqrt{2}=\ket{\uparrow_z\uparrow_z}$, we conclude $\alpha=0$. Only in this case the input state remains unchanged and is an eigenstate. The transformation applied to the three described input states can therefore determine the three phases in the above matrix.\\\\
We now test the three conclusions, $\alpha=\beta=\gamma=0$, against our gate implementation. First and second, we produce the two aforementioned Bell states, $\ket{\Phi^+}$ and $\ket{\Psi^+}$, from the respective input states $\ket{\uparrow_x\uparrow_x}$ and $\ket{\uparrow_x\downarrow_x}$, as described in the main text and reported in Table \ref{tab:bell_states} and Fig. \ref{fig:bell}. Third, we employ the input state $\ket{\uparrow_z\uparrow_z}$, noting that in our protocol the second atom experiences only a photon reflection, nothing that could rotate the atom out of its initial energy eigenstate. We confirm the immutability of the input state by the experimental observation that two reflections of a photon preserve the state with a probability of $(98.6\pm0.6)\%$ (361 detected photon events). From this we infer that the energy eigenstate is indeed an eigenstate under photon reflection.
\subsection*{Reconstructed Density Matrices after the Gate Protocol}
\begin{center}
\begin{figure}
\includegraphics[width=\columnwidth]{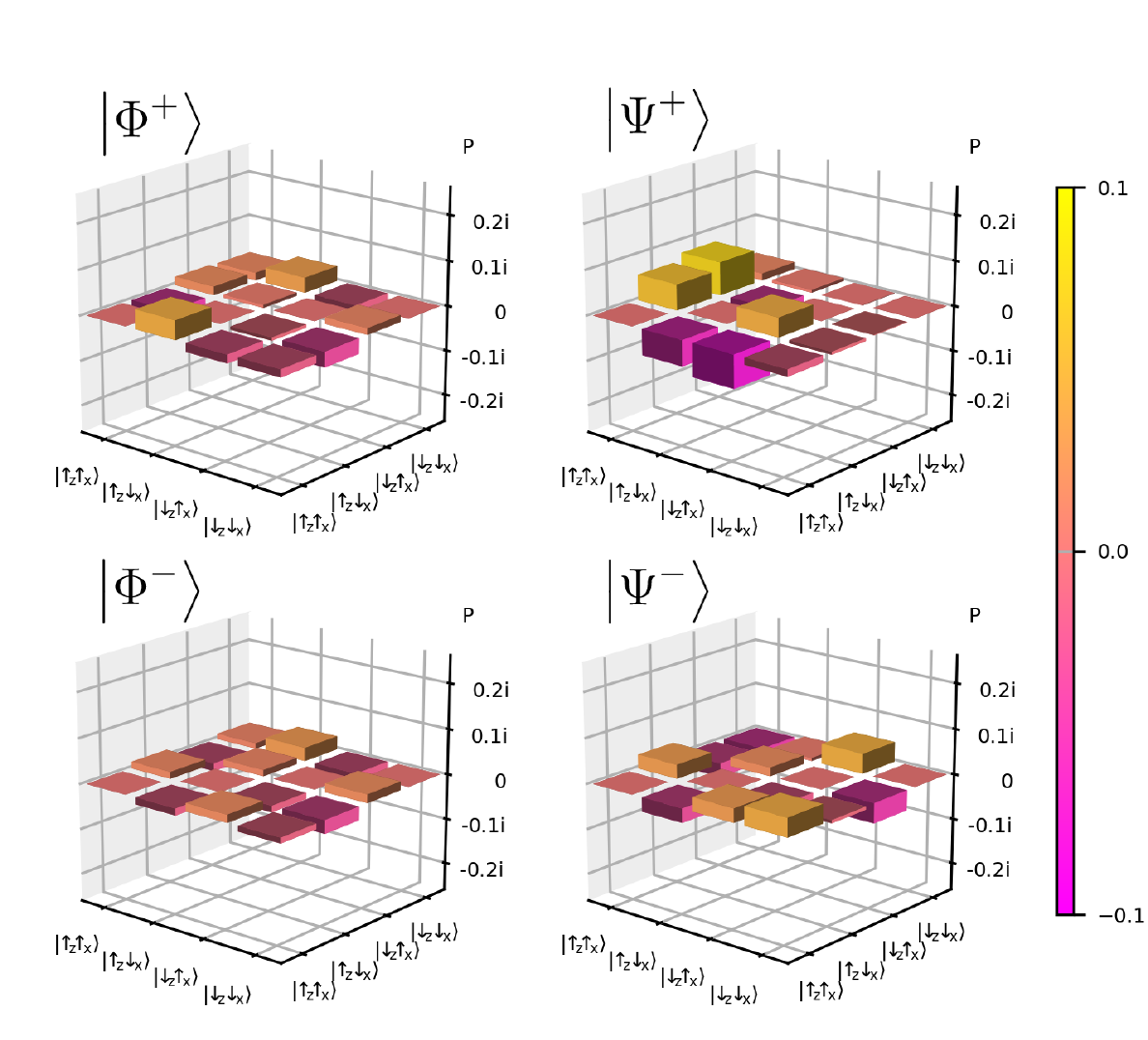}
\caption{\textbf{Density matrices for all four produced Bell states.} Imaginary part of the reconstructed density matrices are shown.}
\label{fig:supp_bell}
\end{figure}
\end{center}

\noindent To analyze the combined quantum state of the two distant qubits after using our gate to create all four Bell states, we perform a full tomography as stated in the main text. There we give the real part of the reconstructed density matrices. For completeness, we show in Fig. \ref{fig:supp_bell} the small residual imaginary contributions after the reconstruction.
\subsection*{Simulating the Effects of Errors}
\noindent We simulate the full experiment in python using the QuTip library \cite{Johansson2012}. The interaction of the weak coherent light state with the two quantum modules is calculated using an input-output theory approach with our cQED parameters \cite{Hacker2019}. Furthermore, the simulation uses separately estimated numbers for relevant experimental parameters including transmission and detection efficiencies, qubit decoherence times, cavity detunings and birefringence, the fluctuations related to the lock of the resonators frequencies, transversal mode matchings, the mapping of the polarizations and detector dark counts. The impact of these imperfections on our fidelity is then extracted from the simulation. Errors are obtained from the fluctuations of the experimental parameters.
As discussed in the main text, a major contribution to the observed infidelity results from the use of weak coherent laser pulses and their residual probabilities of higher photon numbers arriving on either resonator. This leads to an infidelity of $(3.8^{+0.5}_{-0.4})\%$ in the entanglement operation and $(2.3^{+0.3}_{-0.3})\%$ for the truth table. In the latter case, the error is less pronounced as there is the same phase shift of the atom-light state independent of polarization and photon number for an atom in $\ket{\downarrow_z}$ that does not couple to the cavity.
A further limitation stems from the close to $400\mu s$ long coherence times of the atomic states that is largely affected by spurious magnetic field fluctuations and differential trap frequencies. The effect of decoherence on the entanglement fidelity is $(3.2^{+0.3}_{-0.2})\%$ and is 
mainly influenced by the dephasing of the atom in the first module, as it needs to retain coherence throughout the feedback pulses. For the truth table data, the first atom is prepared in an energy eigenstate which greatly suppresses the effect of decoherence on the fidelity to $(0.11^{+0.09}_{-0.03})\%$.

For the whole non-local gate protocol it is important to match the polarizations at the different modules and the detection setup. Small drifts, polarization dependent losses, the limited accuracy of the fiber compensation and cavity birefringence effects pose a limitation on the overlap of these polarizations. From separate measurements we estimate this to have an effect of $(2.7^{+0.4}_{-0.6})\%$ and $(2.2^{+0.3}_{-0.5})\%$ on the fidelity of our protocols.

Furthermore, there are errors of $(2.7^{+1.0}_{-1.0})\%$ and $(2.9^{+1.0}_{-1.0})\%$ related to our state preparation and measurement, which includes drifts in magnetic fields affecting our qubit rotations and thus both the initial states and the measurement bases for the tomography. Frequency fluctuations and drifts between the two resonators -- the phase shift at each photon reflection depends on the detuning between the light and the respective resonator -- cause errors of $(1.9^{+2.8}_{-1.0})\%$ (or $(1.4^{+1.3}_{-0.7})\%$ for the truth table). Lastly, there are errors of $(2.4^{+2.5}_{-1.5})\%$ and $(2.0^{+1.6}_{-1.2})\%$ related to the imperfect transversal matchings between the fiber modes and the resonator modes that lead to an admixture of light that has not interacted with the atom-cavity system. Smaller effects stemming from detector dark counts and the cQED parameters are in the sub-percent range. The remaining difference between the simulated fidelity and the measured values we attribute to additional drifts and a possible underestimation of fluctuations of some of the relevant parameters during the time of the measurement.
Overall, we expect to reduce theses sources of error with measures as using true single photons, implementing faster single qubit rotations and better long-term stability of the relevant frequencies and polarizations.
  
\end{document}